\documentclass[10pt, conference, letterpaper]{IEEEtran}
\usepackage{cite}
\usepackage{amsmath,amssymb,amsfonts}
\usepackage{xfrac} 
\usepackage{graphicx}
\usepackage{xcolor}
\usepackage[mode=match, text-family-to-math = true ,
text-series-to-math = true]{siunitx}
\usepackage[nolist,nohyperlinks]{acronym}
\usepackage{selinput}\SelectInputMappings{adieresis={ä},germandbls={ß}}
\DeclareSIUnit{\bit}{b}
\usepackage{xurl}
\usepackage[font=footnotesize]{caption} 
\usepackage[font=footnotesize]{subcaption}
\DeclareCaptionLabelSeparator{tsep}{\par}
\usepackage{float}
\usepackage{booktabs}
\usepackage{balance} 

\begin{document}
\bstctlcite{IEEEexample:BSTcontrol} 

\begin{acronym}
    \acro{5G}{fifth generation}
    \acro{6G}{sixth generation}
    \acro{AED}{adapted energy detector}
    \acro{AUC}{area under the ROC curve}
    \acro{CU}{central unit}
    \acro{DBSCAN}{density-based spatial clustering of applications with noise}
    \acro{DT}{digital twin}
    \acro{FPR}{false positive rate}
    \acro{JCAS}{joint communications and sensing}
    \acro{LOF}{local outlier factor}
    \acro{ML}{machine learning}
    \acro{NPN}{non-public network}
    \acro{OCSVM}{one-class support vector machine}
    \acro{PT}{physical twin}
    \acro{REM}{radio environment map}
    \acro{ROC}{receiver operating characteristics}
    \acro{RM}{radio map}
    \acro{RSS}{received signal strength}
    \acro{STFT}{short-time Fourier transform}
    \acro{SU}{sensing unit}
    \acro{TPR}{true positive rate}
    \acro{TRP}{total radiated power}
\end{acronym}

\title{Digital Twin of the Radio Environment: A Novel Approach for Anomaly Detection in Wireless Networks}

\author{\IEEEauthorblockN{Anton Krause, Mohd Danish Khursheed, Philipp Schulz, Friedrich Burmeister and Gerhard Fettweis}
\IEEEauthorblockA{Vodafone Chair Mobile Communications Systems, Technische Universit{\"a}t Dresden, Germany\\
Email: \{anton.krause, mohd\_danish.khursheed, philipp.schulz2, friedrich.burmeister, gerhard.fettweis\}@tu-dresden.de}}

\maketitle

\begin{abstract}
The increasing relevance of resilience in wireless connectivity for Industry 4.0 stems from the growing complexity and interconnectivity of industrial systems, where a single point of failure can disrupt the entire network, leading to significant downtime and productivity losses. It is thus essential to constantly monitor the network and identify any anomaly such as a jammer. Hereby, technologies envisioned to be integrated in \acs{6G}, in particular \ac{JCAS} and accurate indoor positioning of transmitters, open up the possibility to build a \ac{DT} of the radio environment. This paper proposes a new approach for anomaly detection in wireless networks enabled by such a \ac{DT} which allows to integrate contextual information on the network in the anomaly detection procedure. The basic approach is thereby to compare expected \acp{RSS} from the \ac{DT} with measurements done by distributed \acp{SU}. Employing simulations, different algorithms are compared regarding their ability to infer from the comparison on the presence or absence of an anomaly, particular a jammer.   Overall, the feasibility of anomaly detection using the proposed approach is demonstrated which integrates in the ongoing research on employing \acp{DT} for comprehensive monitoring of wireless networks.
\end{abstract}

\begin{IEEEkeywords}
Anomaly detection, digital twin (DT), machine learning (ML), resilience
\end{IEEEkeywords}


\section{Introduction}

\acresetall

\Ac{6G} mobile networks will follow the trend of \ac{5G} networks to connect more and more things from the real world, such as machines, sensors, vehicles, etc.~\cite{nguyen20226g}. As the number of wirelessly connected devices increases, so does the potential risk posed by interference from external sources. Imagine a factory that is jammed, no matter whether it happens intentionally or unintentionally (e.g., out-of-band radiation from a device that does not meet the regulations). The disruption of wireless communications could lead to immense costs by production downtime or even to life-threatening situations in scenarios where humans and robots cooperate. Thus, resilience, i.e., the ability to maintain functionality under adverse conditions, is envisioned as one of the main assets of future \ac{6G} mobile networks~\cite{fettweis20226g}.

To enable a network to take countermeasures against an abnormal status, the first step is to detect the anomaly, a topic that has received strong attention from the research community recently. Many recent works that consider jamming on the physical layer, process spectrograms using \ac{ML} to detect whether there is an anomaly. Thereby, both supervised and unsupervised learning approaches are described in the literature. For the supervised learning approach, both normal and abnormal spectrograms are provided to the model in order to learn the classification of them~\cite{xu2022neural, wu2017jamming}. Yet, supervised learning has the shortcoming that types of anomalies not seen in the training phase are probably also not recognized in the operational phase. Thus, unsupervised learning has been examined, either based on autoencoders applied to the spectrogram~\cite{rajendran2018saife, zhou2021radio}, whereby improperly reconstructed parts are considered as an anomaly, or prediction of signals~\cite{tandiya2018deep}, where deviations from the prediction are regarded as an anomaly. Another approach for anomaly detection is to monitor parameters such as the bit error rate and packet error rate, as described for example in~\cite{manesh2019performance} where also supervised learning is used.


However, none of the presented approaches integrates contextual information on the network, e.g., the number and accurate location of active regular transmitters, information which are envisioned to be available in future \ac{6G} networks. Based on the increasing need for resilience in future wireless networks, we present a novel approach for anomaly detection in wireless networks which incorporates such information by building a \ac{DT} of the radio environment. While already widely employed in manufacturing, \acp{DT} of telecommunication systems still have a huge potential to unfold and are therefore in the focus of research from both industry and academia~\cite{ahmadi2021networked, nguyen2021digital, lin20226g}. \Acp{DT} are virtual representations of physical systems which are based on accurate digital models as well as interconnections between the \ac{DT} and its physical world counterpart, the so called \ac{PT}~\cite{kuruvatti2022empowering}. First, the system architecture of the \ac{DT} of the radio environment is introduced. The radio environment as considered in this work is based on the concept of \acp{REM}, which already have been widely described in the literature, for example in~\cite{pesko2014radio}. It includes the transmitters as well as the physical environment (obstacles and their materials, etc.) and propagation characteristics. Based on simulations we then show how anomalies in the radio environment -- particularly jammers -- can be detected using the proposed system together with anomaly detection algorithms. Those algorithms require no prior knowledge on the characteristics of potential jammers but do the inference only from normal operational data. This work provides an initial proof of concept rather than a comprehensive performance evaluation. Therefore, we conclude with an outlook on future research directions.
\section{System Architecture}
\label{sec:system_architecture}

\begin{figure}[t]
    \centering
    \includegraphics[width=\columnwidth]{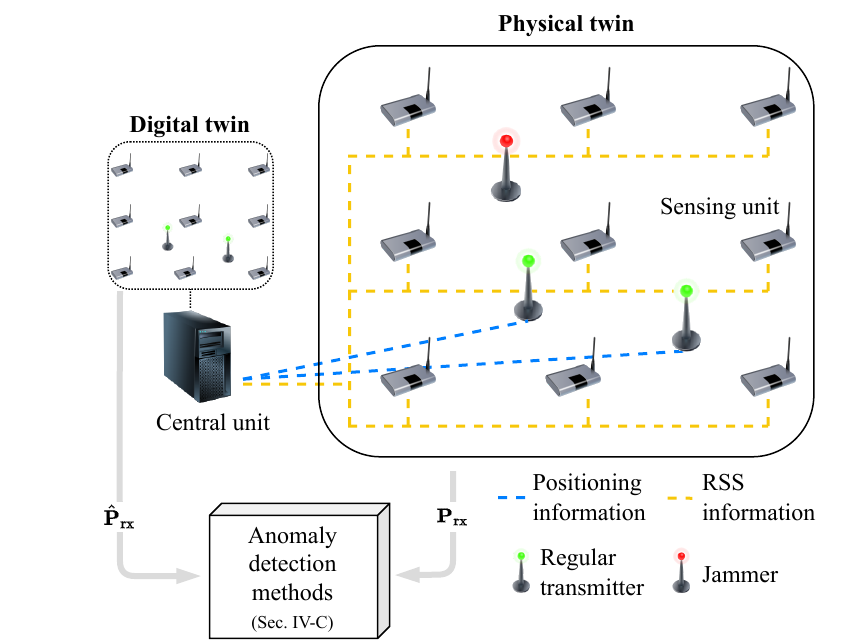}
    \caption{System architecture}
    \label{fig:system_architecture}
\end{figure}

An overview of the system intended for anomaly detection is provided in Fig.~\ref{fig:system_architecture}. The system consists of several \acp{SU} which are distributed over the area to be monitored. Each \ac{SU} measures the \ac{RSS} at its fixed and exactly known location and provides it to a \ac{CU}. The vector of \ac{RSS} measurements is denoted as $\mathbf{P}_\text{rx}$ in which each \ac{SU} is represented by one value (see Section~\ref{sec:system_model} for details). The system is intended for use in a licensed band where the regular transmitters are known. Additionally, the location of each regular transmitter is known by the \ac{CU}. The positioning itself is considered to be provided by the \ac{6G} system and will not be further discussed here apart from the positioning accuracy in Section~\ref{sec:building_the_dt}.

The \ac{CU} has furthermore an accurate database on the physical environment available which can originate for example from the \ac{JCAS} abilities of the \ac{6G} system or from other types of sensors, such as LIDAR or camera. For now, we assume isotropic transmitters, meaning that transmitter orientation has no effect on the modeled radio environment.
Employing the available information about the transmitters, the physical environment database, and a database that contains the electromagnetic properties of different materials, propagation modeling (e.g., ray tracing) can be applied in a regular manner to estimate the expected \ac{RSS} at the locations of the \acp{SU}. The vector of estimated \ac{RSS} values is denoted as $\hat{\mathbf{P}}_\text{rx}$ All the information processed in the \ac{CU} together with the propagation modeling is denoted as \ac{DT} of the radio environment within the scope of this work.

Subsequently, the \ac{CU} compares the expected \ac{RSS} from the \ac{DT} with the \ac{RSS} values that are actually measured by the \acp{SU}. The differences between the measured and the expected \ac{RSS} values for all deployed \ac{SU} serve then as input for the anomaly detection which is explained in detail in Section~\ref{sec:methods}.

Due to its computational complexity, the proposed approach is intended for monitoring areas with limited spatial dimensions and increased resilience requirements (e.g., factories) rather than large-scale and non-critical networks.

\section{System Model}
\label{sec:system_model}

Since resilience is of prominent significance for wireless networks in production environments, we orient our setting towards indoor campus networks, also referred to as \acp{NPN}. The detailed system model used for simulations and data generation is described in this section.

The considered area has a size of $\SI{40}{m} \times \SI{40}{m}$. Obstacles are not modeled to keep the scenario general. A random number $N_\text{reg}$ of regular transmitters is active at random locations $\mathbf{u}_i$ ($i \in \{1, \dots , N_\text{reg}\}$). All transmitters are assumed to use isotropic antennas and transmit with a power ${P_{\text{tx,reg}} = \SI{20}{dBm}}$ at a carrier frequency ${f_c = \SI{3.7}{GHz}}$, whereby the values are oriented on the regulations for \ac{NPN} in Germany \cite{bundesnetzagentur2019}. In the current state of our work, we assume the transmit power to be known exactly by the system and also to be constant, i.e., there is no power control active. Furthermore, there are $N_\text{SU}$ \acp{SU} which measure the \ac{RSS} at fixed locations $\mathbf{v}_j$ ($j \in \{1, \dots , N_\text{SU}\}$) (see Section~\ref{sec:system_architecture}). For the scope of this work the \acp{SU} are arranged in a grid, but other constellations would be conceivable as well. Unless otherwise stated, a grid size of \SI{10}{m} is used, leading to $N_\text{SU} = 25$.

The path loss $L$ is modeled using the log-distance path loss model with log-normal shadowing~\cite{rappaport2002wireless}
\begin{equation}
\label{eq:log_distance_path_loss_plus_noise}
    \resizebox{.9\hsize}{!}{$\displaystyle
    L_{i,j} [\si{dB}] = 10 \; \alpha \log_{10} \left( \frac{d_{i,j}}{\si{m}} \right) + 20 \log_{10} \left( \frac{f_c}{\si{Hz}} \right) + L_0 + X_\sigma,
    $}
\end{equation}
where $d_{i,j}$ denotes the distance between transmitter~$i$ and \ac{SU}~$j$, $L_0$ the path loss offset and $X_\sigma$ the shadowing, which is zero-mean Gaussian distributed with a standard deviation $\sigma$ (in dB). The shadowing is correlated with the covariance matrix $\mathbf{C}$, whereby the entries are given by
\begin{equation}
    \label{eq:cov_matrix_entry}
    [\mathbf{C}]_{k,l} = \sigma^2 \exp\left( -\frac{d(\mathbf{x}_k, \mathbf{x}_l)}{d_\text{cor}} \right)
\end{equation}
for two points $\mathbf{x}_k$ and $\mathbf{x}_l$, separated by the distance $d(\mathbf{x}_k, \mathbf{x}_l)$. Measurements have shown that the correlation in indoor environments is typically limited to small areas~\cite{jalden2007correlation}, thus we use a correlation distance $d_{\text{cor}} = \SI{1}{m}$ in this work.

\begin{table}[h]
\centering
\caption{simulation parameters}
\label{tab:simulation_parameters}
\renewcommand{\arraystretch}{1.2}
\begin{tabular}{ll}
\toprule
Parameter                                               & Value \\
\midrule
Carrier frequency $f_c$                                 & \SI{3.7}{GHz}                                         \\
No. of regular transmitters $N_\text{reg}$              & 10                                                    \\
Transmit power of regular transmitters $P_\text{tx, reg}$                              & \SI{20}{dBm}                                          \\
No. of jammers $N_\text{jam}$                           & 0 or 1                                                \\
Transmit power of jammer $P_\text{tx, jam}$                              & \SI{20}{dBm}                                          \\
Number of sensing units $N_\text{SU}$                   & 25                                                    \\
Path loss exponent  $\alpha$                            & 2                                                     \\
Path loss offset $L_0$                                  & \SI{-147.55}{dB}                                      \\
Positioning error distribution ($x$ and $y$ direction)  & $\mathcal{N} (0, \, \SI{1.33}{m})$                    \\
\bottomrule
\end{tabular}
\end{table}

In addition to the regular transmitters, there may be $N_\text{jam}$ jammers active. For the sake of simplicity, we consider only one jammer in this work, but we expect that our approach will also be able to cope with a higher number of jammers which will be verified in our future work. The jammer (if present) is located at the position $\textbf{w}$ equipped with an isotropic antenna and also transmits at a carrier frequency of \SI{3.7}{GHz}. The transmit power $P_{\text{tx,jam}}$ is fixed to \SI{20}{dBm} as if the jammer would mimic a regular transmitter. An overview of the parameters is given in Table~\ref{tab:simulation_parameters}.

The system model is deliberately chosen simple for the proof of concept presented in this work and to able to draw conclusion which are not drawn to a specific scenario. For our future work, we will evaluate the concept in real-world oriented scenarios (see Section~\ref{sec:future_work}).
\section{Methods}
\label{sec:methods}


The detection of anomalies in the radio environment as introduced in this work bases on the construction of a \ac{DT} and its comparison with real-world measurements. While the building process is described in the first part of this section, the methodology to detect anomalies from the difference between real-world and \ac{DT} measurements is described in the second part.

\subsection{Building the Digital Twin}
\label{sec:building_the_dt}

Various information from different domains allow to build a virtual representation of the radio environment, the so called \ac{DT}. As mentioned in Section~\ref{sec:system_model}, this is basically done by employing propagation modeling for the regular transmitters. For the sake of simplicity, we employ in this work isotropic transmitters. While transmit powers can be measured relatively accurate, radio-based localization in indoor environments is still challenging. Thus, positioning inaccuracies are modeled as follows. According to \cite{behravan2023positioning}, for uplink time difference of arrival localization in FR1 we can expect a positioning error of less than \SI{2.19}{m} in 90\% of the cases. For modeling it is assumed that the error in $x$ and $y$ direction is normally distributed and uncorrelated, leading to a Rayleigh distribution for the magnitude of the error vector. From this, a standard deviation of the error in $x$ and $y$ direction of  \SI{1.02}{m} is derived.

\begin{figure*}
    \centering
    \begin{subfigure}[t]{0.32\textwidth}
        \includegraphics[height=4.5cm]{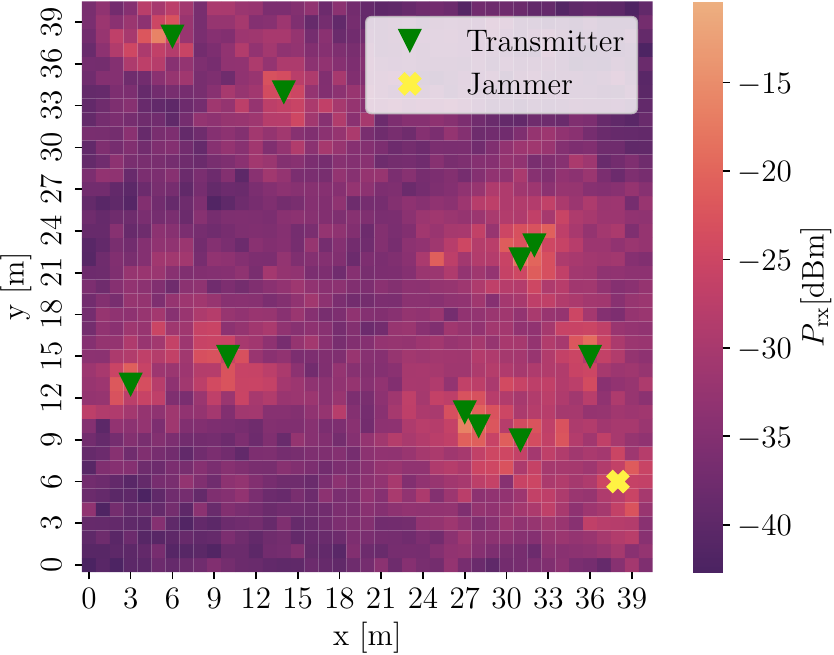}
        \caption{Original radio map}
        \label{fig:orig_rm}
    \end{subfigure}
    \begin{subfigure}[t]{0.32\textwidth}
        \centering
        \includegraphics[height=4.5cm]{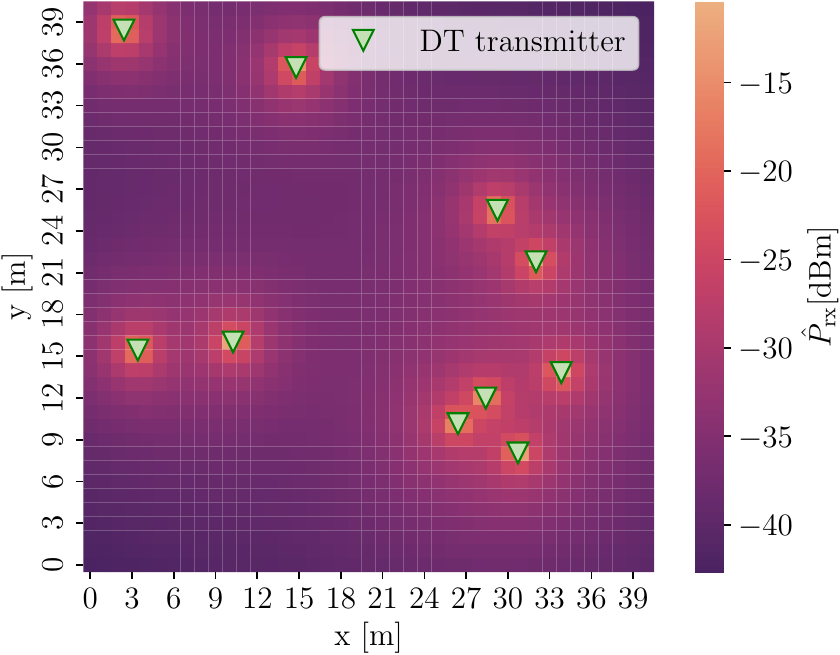}
        \caption{\ac{DT} radio map}
        \label{fig:dt_rm}
    \end{subfigure}
    \begin{subfigure}[t]{0.34\textwidth}
        \centering
        \includegraphics[height=4.5cm]{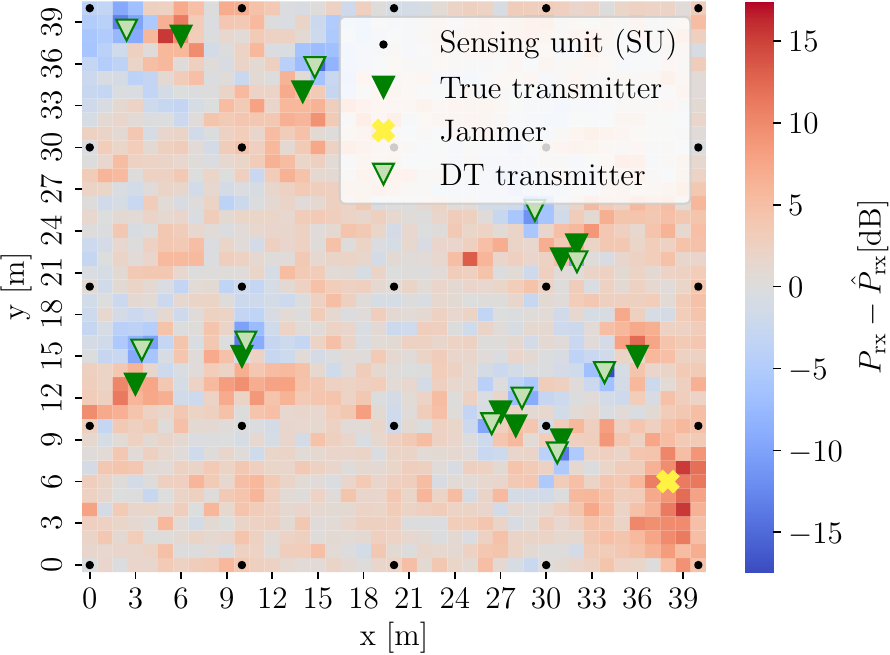}
        \caption{Difference between original and \ac{DT} radio maps}
        \label{fig:diff_rm}
    \end{subfigure}
    \caption{Exemplary radio maps as representation of the radio environment at $\sigma=\SI{4}{dB}$. Measurements are only available at the \ac{SU} positions indicated in (c)}
    \label{fig:rm_example}
\end{figure*}

With the given information on the transmitter, the path loss $\hat{L}_{i,j}$ between transmitter $i$ and \ac{SU} $j$ is estimated using the log-distance path loss model  
\begin{equation}
\label{eq:log_distance_path_loss}
    \hat{L}_{i,j} [\si{dB}] = 10 \; \alpha \log_{10} \left( \frac{\hat{d}_{i,j}}{\si{m}} \right) + 20 \log_{10} \left( \frac{f_c}{\si{Hz}} \right) + L_0
\end{equation}
similar to Eq.~\eqref{eq:log_distance_path_loss_plus_noise}. Because the true distance between the transmitter and the \ac{SU} is unknown, it is replaced by the estimated distance $\hat{d}_{i,j}$ between the estimated transmitter location $\mathbf{\hat{u}}_i$ and the true location of \ac{SU} $j$. The shadowing term $X_\sigma$ is omitted as it is unknown. The difference between the original radio environment and its \ac{DT} is visualized in Fig.~\ref{fig:rm_example}. For a better understanding, not only the \ac{RSS} at the \ac{SU} locations (which are required for anomaly detection) is shown but instead the radio map of the complete area. Fig.~\ref{fig:orig_rm} shows an example radio map with ten regular transmitters and one jammer present. The radio map of the \ac{DT} in Fig.~\ref{fig:dt_rm} incorporates the regular transmitters but neither the jammer nor the random shadowing. The resulting difference is shown in Fig.~\ref{fig:diff_rm}. Big deviations occur close to the positions of the regular transmitters due to localization inaccuracy and at the jammer location, as the jammer is not present in the \ac{DT}. Note, that Fig.~\ref{fig:diff_rm} shows the difference between the original and the \ac{DT} radio map on the whole area, but as input for the anomaly detection problem only the values at the \ac{SU} locations (indicated by the black dots) are used.

\subsection{Fundamentals of the Anomaly Detection}
\label{sec:anomaly_detection_funcamentals}

Given a collection of data samples, an anomaly is one or a group of samples that is rare and differs significantly from the majority of samples in the overall collection~\cite{toshniwal2020overview}. In this work, we consider the presence of a jammer as an anomaly. Jamming can be intentional, i.e., aiming to disturb the operation of the network, but it can also be unintentional. Unintentional jamming might be for example out of band radiation of a device in a neighboring band that does not meet the regulations or an interfering neighboring \ac{NPN}.

The detection of anomalies in the radio environment as proposed in this paper is based on comparing the \ac{RSS} measured by the \acp{SU} at different locations with the \ac{RSS} expected from the \ac{DT} of the radio environment. With the fundamentals given in the previous section, the detection problem in the linear domain can be formulated as follows
\begin{equation}
\label{eq:hypothesis}
    P_{\text{rx},j} [\si{mW}] = 
    \begin{cases}
    \mathcal{H}_0: \; \displaystyle\sum_{i=1}^{N_\text{reg}} \cfrac{P_{\text{tx},i}}{L_{i,j}}    \\[12pt]
    \mathcal{H}_1: \; \displaystyle\sum_{i=1}^{N_\text{reg}} \cfrac{P_{\text{tx},i}}{L_{i,j}} + \cfrac{P_\text{tx, jam}}{L_{\text{jam}, j}}
    \end{cases},
\end{equation}
where $\mathcal{H}_0$ refers to the hypothesis that the received power at receiver $j$ only sums up from the powers received by all regular transmitters. The alternative hypothesis $\mathcal{H}_1$ refers to the hypothesis that there is a jammer present which contributes additional power.

However, the exact path loss $L_{i,j}$ is not known. Thus, to further elaborate the detection problem, the vector $\pmb{\Delta}$ of differences $ \Delta_j $  between the measured \ac{RSS} values $P_{\text{rx},j}$ and the \ac{RSS} values expected from the \ac{DT} $\hat{P}_{\text{rx},j}$ at each \ac{SU} $j$ is defined. The entries are given by
\begin{equation}
\label{eq:rss_diff}
    \resizebox{.9\hsize}{!}{$\displaystyle
    \Delta_j [\si{dB}] = P_{\text{rx},j} [\si{dBm}] - \hat{P}_{\text{rx},j} [\si{dBm}] \; , \quad j \in \{ 1, \dots , N_\text{SU} \}.
    $}
\end{equation}
This vector serves as input for the anomaly detection algorithms. For the detection, each time the measurements at a single time instance are regarded. Based on Eq.~\eqref{eq:rss_diff} and under the assumption, that the transmit power $P_\text{tx,reg}$ of regular transmitters is exactly known, the detection problem in Eq.~\eqref{eq:hypothesis} can be reformulated to
\begin{equation}
\label{eq:hypothesis_diff}
    \resizebox{.9\hsize}{!}{$\displaystyle
    \Delta_j [\si{dB}] = 
    \begin{cases}
    \mathcal{H}_0: \; 10 \log_{10} \left( \displaystyle\sum_{i=1}^{N_\text{reg}} \cfrac{P_{\text{tx},i}}{L_{i,j}} \right) - 10 \log_{10} \left( \displaystyle\sum_{i=1}^{N_\text{reg}} \cfrac{P_{\text{tx},i}}{\hat{L}_{i,j}} \right)   \\[16pt]
    \mathcal{H}_1: \; 10 \log_{10} \left( \displaystyle\sum_{i=1}^{N_\text{reg}} \cfrac{P_{\text{tx},i}}{L_{i,j}} + \cfrac{P_\text{tx, jam}}{L_{\text{jam}, j}} \right) - 10 \log_{10} \left( \displaystyle\sum_{i=1}^{N_\text{reg}} \cfrac{P_{\text{tx},i}}{\hat{L}_{i,j}} \right)
    \end{cases}
    $},
\end{equation}
with powers and losses given in linear scale. This means that the anomaly detection problem boils down to the decision whether the differences in $\pmb{\Delta}$ can be purely explained by inaccuracies in path loss modeling (due to shadowing and localization inaccuracy) or whether there is a jammer present.

\subsection{Anomaly Detection Methods}
\label{sec:unsupervised_learning}

The procedure to implement the anomaly detection system is intended as follows. Based on models of the physical environment and propagation models, the \ac{DT} of the radio environment is established first. In the next phase, denoted as training phase, training samples are collected during normal operation. Those samples are assumed to be free from anomalies and are used to train the algorithms described later on in this section. The task is thereby to generalize the statistics of normal data (i.e., of the modeling error) and to be able to identify samples as anomaly that seem to stem from a different generation process. This approach allows to identify jammer even without any prior knowledge their characteristics. Thus, the general problem is also often referred to as novelty detection. When the training is finished, the system starts its regular operation to detect anomalies.

In the following, the three applied algorithms are introduced. \Ac{OCSVM} and \ac{LOF} are well-known unsupervised \ac{ML} algorithms for anomaly (outlier) detection, while the \ac{AED} is inspired by the energy detector for signal detection tasks. For \ac{OCSVM} and \ac{LOF}, the \textit{scikit} library~\cite{scikit-learn} implementations are used, whereas \ac{AED} is self implemented.

\subsubsection{\Acf{AED}} The concept of energy detectors is well known and has already been widely applied, for example in cognitive radio~\cite{atapattu2014energy}. The task thereby is to detect an unknown signal in the presence of noise. Given the decision problem in Eq.~\eqref{eq:hypothesis_diff}, the problem can be approached from a similar perspective. Either $\Delta_j$ originates only from random model inaccuracies in case $\mathcal{H}_0$ or there is a jamming signal present which contributes additional received power. For a perfect \ac{DT} (i.e., $L_{i,j} = \hat{L}_{i,j}$) we could expect either ${\Delta_j = 0}$ in case $\mathcal{H}_0$ or 
\begin{equation}
    \Delta_j [\si{dB}] = 10 \log_{10} \underbrace{ \left( \cfrac{\sum_{i=1}^{N_\text{reg}} \frac{P_{\text{tx},i}}{L_{i,j}} + \frac{P_\text{tx, jam}}{L_{\text{jam}, j}}}{\sum_{i=1}^{N_\text{reg}} \frac{P_{\text{tx},i}}{\hat{L}_{i,j}}}   \right) }_{>1} > 0
\end{equation}
in case $\mathcal{H}_1$. Due to the previously mentioned random and unknown inaccuracies in the \ac{DT}, $\Delta_j$ is randomly distributed even in case no jammer is present. The presence of random log-normal components in the path loss, resulting from shadowing, prevents the existence of a straightforward mathematical expression for the distribution of $\Delta_j$~\cite{schwartz1982on}. Still, the findings justify  the following approach
\begin{equation}
    \overline{\Delta} = \frac{1}{N_\text{SU}} \sum_{j=1}^{N_\text{SU}} \Delta_j \begin{array}{l}
        \geq \Delta_\text{th} \; : \; \text{anomaly} \\
        < \Delta_\text{th}  \; : \; \text{no anomaly}
    \end{array}
\end{equation}
with the threshold $\Delta_\text{th}$. It is derived from the statistics of the training data, e.g., the 90th percentile of $\overline{\Delta}$ in the training data.


\subsubsection{\Acf{OCSVM}} The approach of \ac{OCSVM} is to identify a hypersphere that encloses all (or most) of the data points~\cite{noumir2012on}. This hypersphere is fitted during the training phase. In the test phase, samples are categorized as normal if they fall within the hypersphere or as anomaly otherwise. The default parameters of the \ac{OCSVM} implementation remain untouched.

\subsubsection{\Acf{LOF}} Instead of a binary decision whether a sample is an anomaly or not, \ac{LOF} assigns each sample a score that can be interpreted as the outlier degree of the given sample. This score is also referred to as \ac{LOF} and it is calculated by comparing the density around the point under consideration with the density of points in the neighborhood. The user-specified parameter $k$ thereby defines how many of the closest points belong to the neighborhood of one point~\cite{breunig2000lof}. 

In the scope of this work, $k$ is set to 100 to compensate for the sparsity of the data set due to the high dimensionality. Furthermore, the parameter \textit{novelty} is set to \textit{True} to ensure that the density is calculated based only on the training data. The threshold score about which an outlier is identified as anomaly is varied during the \ac{ROC} analysis.
\section{Results}
\label{sec:results}

The performance of the algorithms under study for anomaly detection as well as the influence of the shadowing level is examined in this section.

To evaluate the performance, a data set is created according to the system model presented in Section~\ref{sec:system_model} containing \num{20000} samples overall. Thereby, \num{10000} normal (i.e., not jammed) samples are used for training and another \num{10000} samples for testing, whereby now half of the samples are anomalies and the other half are normal samples. Different data sets are created to investigate various shadowing levels. Unless otherwise stated in the explanations of the specific algorithms in Section~\ref{sec:unsupervised_learning}, no hyperparameters of the algorithms are tuned, as this would require the evaluation on the test set and thereby mean a leakage of test data (i.e., knowledge of anomalies) into the training phase of the algorithms. After the training phase, the models are evaluated on the test set by means of the metrics presented in the next paragraph. This procedure is repeated three times to improve the reliability of the results. 

The detection is regarded as binary classification problem, where we assign the \textit{positive} class to the anomalies, and the \textit{negative} class are the normal samples. To compare the performance of different algorithms, \ac{ROC} curves are employed. The \ac{ROC} curve depicts the tradeoff between the \ac{TPR} and the \ac{FPR} which are defined in Eq.~\eqref{eq:tpr_fpr}~\cite{fawcett2006roc}.

\begin{equation}
    \label{eq:tpr_fpr}
    \text{TPR} = \frac{\text{TP}}{\text{TP} + \text{FN}} \qquad \text{FPR} = \frac{\text{FP}}{\text{TN} + \text{FP}}
\end{equation}

To create the \ac{ROC}, the decision threshold of the algorithms is varied. For the specific algorithms, this means: for \ac{AED} the threshold $\Delta_\text{th}$, for \ac{OCSVM} the radius of the hypersphere and for \ac{LOF} the score which defines an outlier is varied. Through the parameters, it can be controlled whether algorithms shall act more 'conservative', i.e., identify anomalies only with high confidence at the cost of many missed detections, or more 'liberal', i.e., detecting most of the anomalies at the cost of many false alarms. Additionally, the \ac{ROC} is characterized by the property that it does not change if the ratio between normal and abnormal samples changes~\cite{fawcett2006roc}. Thus, the evaluation can be done without assuming a specific jamming probability. 


\begin{figure}[t]
    \centering
    \begin{subfigure}[t]{0.52\columnwidth}
        \centering
        \includegraphics[height=4.4cm]{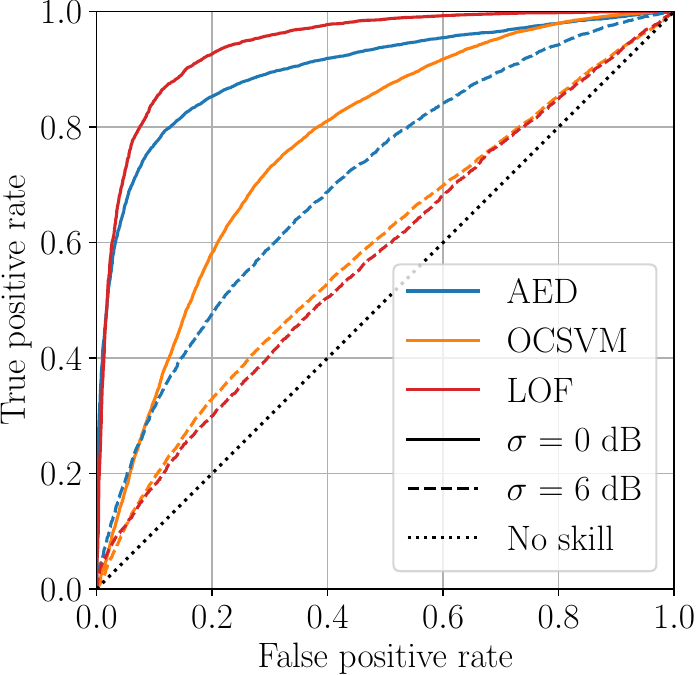}
        \vspace{-0.35cm}
        \caption{Different shadowing levels at a grid size of \SI{10}{m}}
        \label{fig:roc-noise}
    \end{subfigure}
    \begin{subfigure}[t]{0.46\columnwidth}
        \centering
        \includegraphics[height=4.35cm]{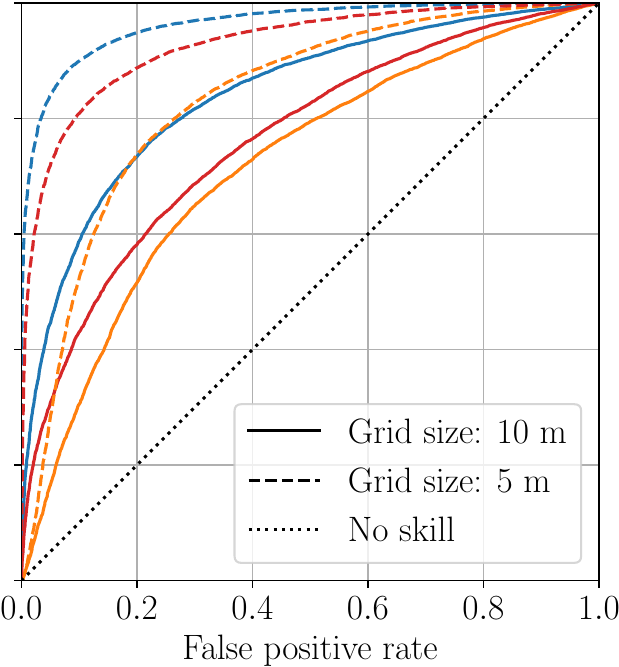}
        \vspace{-0.35cm}
        \caption{Different grid sizes at a shadowing level of $\sigma = \SI{2}{dB}$}
        \label{fig:roc-grid}
    \end{subfigure}
    \caption{\Ac{ROC} curves for the three algorithms}
    \label{fig:roc}
\end{figure}

The \ac{ROC} curves of the three discussed algorithms (see Section~\ref{sec:unsupervised_learning}) are shown in Fig.~\ref{fig:roc-noise} for the shadowing levels $\sigma = \SI{0}{dB}$ and $\sigma = \SI{6}{dB}$. A detector without skill, i.e., which just performs random guesses on whether there is an anomaly, would result in the black dotted line. The better an algorithm performs, the more its \ac{ROC} curve tends towards the upper left corner. Thus, one can conclude that \ac{LOF} outperforms the other two algorithms at $\sigma = \SI{0}{dB}$, whereas \ac{AED} shows the best performance at $\sigma = \SI{6}{dB}$.

Further analysis indicate that the density of the \acp{SU} also has a strong impact on the anomaly detection performance. Reducing the grid size to \SI{5}{m} (i.e., $N_\text{SU} = 81$) significantly improves the performance as shown in Fig.~\ref{fig:roc-grid} exemplarily for $\sigma = \SI{2}{dB}$. However, there is a tradeoff between increased detection performance on the one hand and increased hardware costs and computational complexity on the other hand, which has to be considered for practical application.

\begin{figure}[t]
    \centering
    \includegraphics[width=0.7\linewidth]{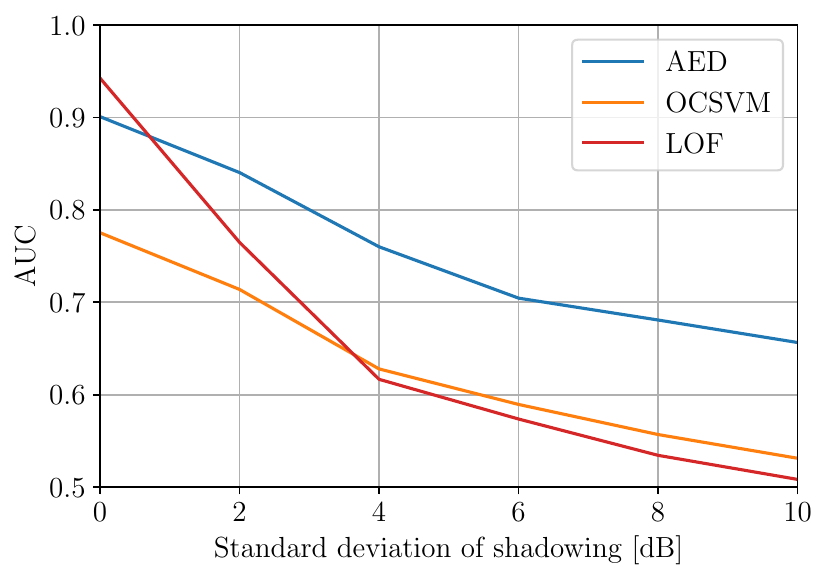}
    \caption{\Ac{AUC} scores for different shadowing levels at a grid size of \SI{10}{m}}
    \label{fig:auc}
\end{figure}

To allow a numeric comparison between the \ac{ROC} curves of different algorithms, the \acf{AUC} was established. It is obtained by integrating the area under a specific \ac{ROC} curve. Visualizing the \ac{AUC} values at different shadowing levels as in Fig.~\ref{fig:auc}, one can see that the performance of all three algorithms degrades with an increasing shadowing level as it can be expected. Apart from $\sigma = \SI{0}{dB}$, where \ac{LOF} has the highest \ac{AUC} score, \ac{AED} is the best-performing algorithm at all other shadowing levels. This is achieved by exploiting the general knowledge that each jammer, no matter which type, has to transmit a signal and thereby increases the \ac{RSS}.  At high noise levels, the \ac{AUC} scores of \ac{LOF} and \ac{OCSVM} approach the value of 0.5, i.e., there is not a big performance difference compared to a detector with no skill, whereas \ac{AED} still has an \ac{AUC} greater than 0.65.

Overall, it has been demonstrated in this section that anomalies in wireless networks -- in particular jammers --  can be detected employing the introduced \ac{DT} approach and unsupervised learning. Nevertheless, as can be concluded from Fig.~\ref{fig:auc}, an accurate modeling of the radio environment is required to achieve a reliable detection. Otherwise, it is challenging for the algorithms to infer whether high deviations between measured and expected \ac{RSS} originate from model errors or from jammers.
\section{Future Work}
\label{sec:future_work}


This paper presents a system architecture and an initial proof of concept on how a \ac{DT} can be used for anomaly detection in wireless networks. Nevertheless, significant steps need to be taken before the concept can be successfully translated into practical application. Thus, we plan to address the following steps in our future research:

\subsubsection{Detection methodology} For the initial proof of concept presented in this work, basic algorithms for unsupervised learning have been applied. For the future work, we plan to evaluate the performance of more sophisticated algorithms and improved preprocessing (e.g., dimensionality reduction).
\subsubsection{Ray tracing} To achieve a high modeling accuracy for the \ac{DT}, the log-distance path loss model shall be replaced with ray tracing to achieve a more accurate model of the path loss. At first glance, the high computational complexity of ray tracing might limit the real-time capability of the proposed system. However, in the literature it has already been demonstrated that ray-tracing might be replaced by \ac{ML} with a comparable accuracy but a drastically reduced computational complexity~\cite{bakirtzis2022deepray}.
\subsubsection{Antenna pattern} In our future work, we will integrate different antenna patterns and hence also the orientation of the transmitter.
\subsubsection{Real-world deployment} Employing ray tracing to build the \ac{DT} enables us to test the proposed concept also in real-world environments. From this, we expect interesting insights, e.g., about the required modeling accuracy (geometry, materials, ray tracing) for deploying an effective anomaly detection system for the radio environment.
\section{Conclusion}
\label{sec:conclusion}

Resilience of the wireless network is key for many of its applications. Therefore, it is essential to constantly monitor the network and identify anomalies to enable a quick reaction to (potentially) critical situations. This work contributes by presenting a novel approach for anomaly detection in wireless networks by employing a \ac{DT} of the radio environment, an approach that allows to incorporate contextual information in the anomaly detection procedure. Furthermore, no prior knowledge on jammer characteristics is required. A system architecture is introduced and the suitability of the approach for anomaly detection is demonstrated based on simulations.
This work serves an initial study of the concept and proposes numerous directions for future studies, e.g., including ray tracing for the \ac{DT} to be able to cope with real-world scenarios instead of statistical models.

\section*{Acknowledgment} \small 
This work was supported by the Federal Ministry of Education and Research, Germany (BMBF) as part of the projects {"6G-CampuSens"} under contract 16KISK207, “Industrial Radio Lab Germany (IRLG)” under contract 16KIS1010K, and "6G-life” under contract 16KISK001K. The authors alone are responsible for the content of the paper.

\balance
\bibliographystyle{IEEEtran}
\bibliography{database}

\end{document}